# Performance Evaluation of Linear Regression Algorithm in Cluster Environment

Cinantya Paramita[1], Fauzi Adi Rafrastara[2], Usman Sudibyo[3], R.I.W. Agung Wibowo
Department of Information Engineering
Faculty of Computer Science
Universitas Dian Nuswantoro, Semarang, Indonesia
cinantya.paramita@dsn.dinus.ac.id [1], fauziadi@dsn.dinus.ac.id [2],
usman.sudibyo@dsn.dinus.ac.id[3], 111201710079@mhs.dinus.ac.id

*Abstract*—Cluster computing was introduced to replace the superiority of super computers. Cluster computing is able to overcome the problems that cannot be effectively dealt with supercomputers. In this paper, we are going to evaluate the performance of cluster computing by executing one of data mining techniques in the cluster environment. The experiment will attempt to predict the flight delay by using linear regression algorithm with apache spark as a framework for cluster computing. The result shows that, by involving 5 PC's in cluster environment with equal specifications can increase the performance of computation up to 39.76% compared to the standalone one. Attaching more nodes to the cluster can make the process become faster significantly.

## I. INTRODUCTION

High Performance Computing (HPC) becomes one of hot topics in the recent years. Either one of the most popular HPC products in the world is called supercomputer. However, traditional supercomputer is no longer dominant in the area of computing and its availability has changed dramatically because of its high cost and low accessibility factors [1][2]. Thus, it is required the HPC system with low cost and high accessibility. Rajak [3] explains that there are 3 types of modern HPC which has better implementation than the traditional supercomputer in term of accessibility and cost. Those are: grid, cloud, and cluster computing. Grid computing is good for resource balancing, access to additional storage, and reliability. Cloud computing has the benefits in super computing power, high resource availability, virtualization, crash recovery, and flexibility. Whereas cluster computing has some advantages, especially in term of manageability, single system image (SSI), and high availability. Not all computation is suitable for all of those 3 solutions, i.e. grid and cloud computing are overkill to solve the simple data mining problems. So it is required the smaller scale of HPC to solve the simpler problem, and it can be overcome effectively by using cluster computing[1][3][4].

Data mining commonly is processed in a standalone computer. However, for the large dataset, mining the data in a standalone PC can take several seconds, minutes, hours, even days, depend on the hardware specifications. A solution is needed to make mining process become more effective, especially in term of processing time [4][5].

In this paper, we will implement data mining algorithm in a cluster environment to accelerate the mining process. We are going to analyze the performance of cluster computing by comparing with various numbers of nodes and the standalone one as well. We will evaluate the performance one by one so that we can see the significance of cluster computing for data mining processing. Case study of this research is predicting the flight delay by using linear regression algorithm. All the simulations will be conducted in the virtual environment.

This paper consists of six sections. Rest of this paper is organized as follows. The ideas, terminologies, and related researches that conducted by other researchers is discussed in section II. The methodology that used in the experiment is explained in section III. Section IV consists of experiment process and result. In this section, Linear Regression algorithm is executed in the both standalone and cluster environment. Conclusion of this research is presented in the section V, followed by the future work in section VI.

## II. LITERATURE REVIEW

Cluster computing is a part of Superior Computing or High Performance Computing (HPC) [3]. Cluster computing is also known as a part of distributed or parallel processing system [4]. It consists of some interconnected individual computers, through Local Area Network (LAN) [6]. Those interconnected computers are running together as a single integrated source. It has some benefits, such as in term of performance improvement, high availability, cost reduction, and manageability [4][7][8].

According to [4], there are three types of cluster computing, those are High Performance Computing Cluster (HPC Cluster), High Availability Cluster, and Load Balancing Cluster. In this paper, cluster computing's type that are going to be implemented is High Performance Computing Cluster (HPC Cluster). The goal is to accelerate the prediction on flight delay problem.

Flight delay is defined when a carrier lands or takes off latter than its scheduled time for arrival or departure. This phenomenon is becoming common and happened frequently in all over the world. Around 20% carriers have more than 15 minutes delay [9].

PySpark is built to provide users the python API in Spark environment. Python is the most popular programming language or tool, especially for the analytics, data science, and machine learning [10][11]. On the other hand, Apache Spark [12] is a unified analytics engine for large scale data processing. Spark is also big data framework that widely used by both academia and industry. Spark has at least four advantages for performing data analytics [12][13], those are:

- Speed: spark is 100x faster than Hadoop in memory, or 10x on disk. Apache Spark has an Advanced Directed Acyclic Graph (DAG) execution engine to support in-memory computing and acyclic data flow.
- Ease of Use: spark supports many popular programming languages, such as: java, scala, python, and R.
- Generality: spark has huge libraries that can be used to perform many activities, including SQL and DataFrames, Spark Streaming, GraphX, and MLib for machine learning.
- Runs Everywhere: spark can run on top of Mesos, Hadoop, and also in the cloud and standalone. Spark also has the ability to access various data sources, such as: Cassandra, HBase, S3, and HDFS.

On-time flight is important for many parties. Passengers want to arrive several hours earlier or on-time to the destination for their business or appointment. Flight delay can bring the uncertainty for the passenger, lost the time, and sometime it increases the trip cost. On the other hand, the airline companies are also charged for the penalties, fines and additional operational costs for the crew and aircraft retention in airports. As a conclusion, flight delay have some negative impacts, especially on economic aspect, for the passengers, airline companies, and also airports [14].

In this research, the flight delay prediction is conducted by using linear regression method, and it is executed in cluster environment. The result will be compared with the performance when executed in standalone mode. A lot of research papers discussed regarding flight delay prediction using some data mining techniques, such as decision tree [15][16][17], random forest [16][18][17], AdaBoost [16], KNN [16], Naïve Bayes [19], C4.5 [19], Linear Regression [19][17], Gradient Boosting Classifier [20], Deep Learning [21], etc.

Linear Regression is chosen because its popularity to predict something and also become one of the most common tools for analytical needs. It is a well-known algorithm that theoretically logic, appropriate in most applications, and easy to implement [22][23][5]. In this paper, Linear Regression is used to predict the flight delay. As a future work, we will use other methods for the same case study, and compare it with the performance of Linear Regression in cluster environment.

### III. METHODOLOGY

The experiment that conducted in this research has the following details:
- Algorithm: Linear Regression
- Dataset: Flight (2702218 records, with 70% for training and 30% or testing)
- Number of nodes: 1-5 (using virtual machine)
- A computer with specification: Intel Xeon E5620 Processor (Cores: 4, Threads: 8), 16 GB RAM, DDR5 1024 MB VGA, 3TB HDD, Windows 10 OS.

This simulation was performed on top of virtual machine (VirtualBox). All the nodes created here have equal specifications, such as: single CPU Core, 2 GB RAM, Windows 8.1, and PySpark as a cluster computing platform. The dataset was already preprocessed, so we could use it directly.

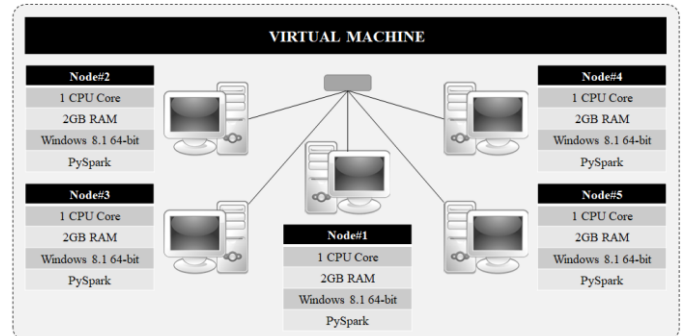

*Figure 1: 5 Virtual PCs created in Virtual Machine with equal specifications*

Apache Spark was used in this research, since it is more powerful than the competitors for the cluster and distributed computing. Spark is 10x faster than Hadoop. Spark is platform to do either batch or stream computing with in memory computing capability. PySpark is a combination of Python and Spark, so that we can use all python's library in Spark environment. Such combination makes both python and spark become more powerful and very useful, especially in data science technology.

### IV. EXPERIMENT & RESULT

We performed the experiment in 5 environments. Those are:
- Standalone: it involves a single PC (in virtual machine) with 2GB RAM. No other PCs connected to this standalone PC.
- Cluster_1: it involves two PCs (in virtual machine) in which 1 PC works as a master node and another PC works as a worker node. In addition, each PC has 2GB RAM.
- Cluster_2: it involves three PCs (in virtual machine), each acquires 2GB RAM. 1 PC works as master node and 2 other PCs work as worker nodes.
- Cluster_3: it involves four PCs (in virtual machine), each has 2GB RAM. There are 3 workers and 1 master node on this cluster.
- Cluster_4: it involves five PCs (in virtual machine) in which 4 PCs have the role of workers and 1 PC for master node.

For the first experiment, we created the first node (Node_1) and run the algorithm trough Jupyter Notebook. This first attempt was conducted in standalone mode. The result was, it took 137.4 seconds to complete the task *(Table 1)*.

Regarding the accuracy of the algorithm, we measured the error prediction by using Root Mean Squared Error (RMSE). RMSE is used to calculate the differences between values that predicted by the algorithm and the values observed [24]. It shows that linear regression algorithm has error prediction around 13,149 minutes for the flight delay dataset. However, this research is not focusing on this case. The accuracy of this algorithm remain the same, even when we conduct the experiment in the cluster environment with more and more nodes attached. In this paper, we are going to highlight the performance of each experiment in term of time consumption, since cluster computing can help us to accelerate the performance speed of data mining computation.

On the next experiment, we started to create a cluster environment with 1 master node and 1 worker node. It resulted a processing time that slightly better than the standalone one (around 123.9 seconds). Faster performance can be acquired when more than 1 worker nodes are attached.

| Environment | Computational Time (in seconds) | | | | | Average |
|---|---|---|---|---|---|---|
| | Experiment | | | | | |
| | 1 | 2 | 3 | 4 | 5 | |
| Standalone (2GB RAM) | 137.4 | 128.6 | 130.7 | 126.7 | 131.7 | 131.02 |
| Cluster_1 (1 worker) - @2GB | 123.9 | 124.1 | 123.7 | 122.6 | 123.3 | 123.52 |
| Cluster_2 (2 worker) - @2GB | 100.4 | 97.5 | 97.6 | 95.3 | 99.5 | 98.06 |
| Cluster_3 (3 worker) - @2GB | 87.1 | 83.5 | 87.2 | 85.5 | 83.7 | 85.4 |
| Cluster_4 (4 worker) - @2GB | 82.9 | 74.1 | 81.2 | 78.9 | 77.5 | 78.92 |

**Table 1: Computation Performance between Standalone and Cluster environment.**

On the third experiment, we used 1 master node and 2 worker nodes. The processing time became faster than the first and second experiment, with 100.4 seconds to complete the task.

The next step was running the algorithm on a master node which supported with 3 worker nodes. We obtained better computational performance compared to 3 experiments in advance. Here, we got 87.1 seconds.

Finally, we used 1 master and 4 worker nodes to increase the speed of computation. It was proofed by getting 82.9 seconds to perform delay prediction on the flight dataset.

We then repeated each experiment 5 times to gain the average time for all computation environments. The average computational time for the standalone, cluster_1, cluster_2, cluster_3, and cluster_4 environments are 131.02, 123.52, 98.06, 85.4, and 78.92 respectively *(Table 1)*.

According to the experiment's result captured in Table 1, by using cluster computing, the computational time can be decreased up to 39.76%. The more computers attached to the cluster, the faster computing time.

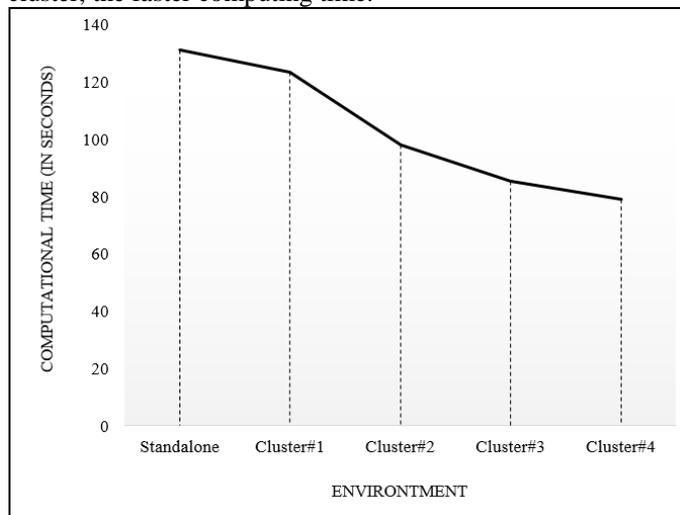

**Figure 2: Computing time has continued to go down when more computers are connected**

These experiments show us that cluster computing is a good alternative to increase the computation performance, especially in data mining field. It is also a cheap way to accelerate the processing speed, thanks to its scalability. We only need the commodity hardware that probably no longer being used by others. Once we found unused PCs, we can attach them directly to our cluster environment so that the computing performance become faster and faster. Better specification of the attached PCs can also affect the computation performance of cluster as well.

V. CONCLUSION

This research was conducted to evaluate the performance of linear regression algorithm to process the data in cluster environment. Linear Regression here was used to predict the flight delay, in which the dataset is publicly available on the internet. We performed the simulation in a virtual environment. The $1^{st}$ experiment was running the algorithm in a virtual standalone PC. On the $2^{nd}$ experiment, we started to develop cluster environment, but consist of 1 master and 1 worker node. The $3^{rd}$ experiment was simulated on 1 master and 2 worker nodes. The $4^{th}$ experiment, the simulation involved 1 master and 3 worker nodes. Lastly, on the $5^{th}$ experiment, we use 5 virtual PCs with 4 PCs as worker and 1 PC as master node.

The performance of this algorithm became faster and faster when more PC attached. The computational time can be decreased up to 39.76% by involving 5 PCs with equal specifications on the cluster. But indeed by attaching more PCs with better specifications can affect the computing performance as well.

VI. FUTURE WORK

As a future work, we will perform the flight delay prediction using other machine learning methods in which implemented on the cluster environment. Those algorithms finally will be compared with the performance of Linear Regression that already conducted in this paper to find the most suitable algorithm for the case of flight delay prediction.